\documentclass[11pt]{article}
\usepackage{epsfig}

\newcommand{\bfm}[1]{\mbox{\boldmath$#1$}}

\catcode`@=11

\fussy
\def\section{\@startsection {section}{1}{\z@}{-3.5ex plus -1ex minus
    -.2ex}{2.3ex plus .2ex}{\bf }}
\def\subsection{\@startsection{subsection}{2}{\z@}{-3.25ex plus -1ex minus
   -.2ex}{1.5ex plus .2ex}{\it }}

\def\@makefnmark{{$\!^{\@thefnmark}$}}

\renewenvironment{thebibliography}[1]
	{\begin{list}{\arabic{enumi}.}
	{\usecounter{enumi}\setlength{\parsep}{0pt}
	 \setlength{\itemsep}{0pt}
         \settowidth
	{\labelwidth}{#1.}\sloppy}}{\end{list}}

\newcounter{arabiclistc}

\def\@citex[#1]#2{\if@filesw\immediate\write\@auxout
	{\string\citation{#2}}\fi
\def\@citea{}\@cite{\@for\@citeb:=#2\do
	{\@citea\def\@citea{,}\@ifundefined
	{b@\@citeb}{{\bf ?}\@warning
	{Citation `\@citeb' on page \thepage \space undefined}}
	{\csname b@\@citeb\endcsname}}}{#1}}

\newif\if@cghi
\def\cite{\@cghitrue\@ifnextchar [{\@tempswatrue
	\@citex}{\@tempswafalse\@citex[]}}
\def\citelow{\@cghifalse\@ifnextchar [{\@tempswatrue
	\@citex}{\@tempswafalse\@citex[]}}
\def\@cite#1#2{{$\!^{#1}$\if@tempswa\typeout
	{IJCGA warning: optional citation argument
	ignored: `#2'} \fi}}

\def\baselinestretch{1.0}
\ifx\selectfont\undefined
\@normalsize\else\let\glb@currsize=\relax\selectfont
\fi

\ifx\selectfont\undefined
\def\@singlespacing{%
\def\baselinestretch{1}\ifx\@currsize\normalsize\@normalsize\else\@currsize\fi%
}
\else
\def\@singlespacing{\def\baselinestretch{1}\let\glb@currsize=\relax\selectfont}
\fi

\long\def\@makecaption#1#2{
   \vskip 10pt
   \setbox\@tempboxa\hbox{\footnotesize #1: #2}
   \ifdim \wd\@tempboxa >\hsize   
       \leftskip 0pt plus 1fil
       \rightskip 0pt plus -1fil
       \parfillskip 0pt plus 2fil
       \footnotesize #1: #2\par   
     \else                        
       \hbox to\hsize{\hfil\box\@tempboxa\hfil}
   \fi}

\catcode`@=12

\begin{document}
\title{\large \bf  THE MASS OF $\bfm \eta_b$\thanks{Talk given at the 44th Rencontres de Moriond }}

\author{\normalsize   A. PENIN$^{a,b}$ \\[2mm]
\small \it $^{a}$Department of Physics,
  University Of Alberta Edmonton, AB T6G 2J1, Canada\\
\small \it  $^{b}$Institute for Nuclear Research,
    Russian Academy of Sciences
  119899 Moscow, Russia}
\date{}

\maketitle\abstract{
In this paper we briefly review the advances and problems in the QCD theory
of the $\eta_b$ mass.}

\section{Introduction}

The properties of the $\Upsilon$ mesons, the bottom quark-antiquark spin-one
bound states, are measured experimentally with great precision, and recent
theoretical analysis of the $\Upsilon$ family based on high-order perturbative
calculations resulted in determinations of the bottom-quark mass $m_b$ with
unprecedent accuracy.\cite{KPP,PenSte}
At the same time  the spin-zero $\eta_b$ meson
remained  elusive despite  dediacted experimental searches.
Only recently a signal of the $\eta_b$
has been observed by Babar collaboration in the radiative decays of the
excited $\Upsilon$ states.\cite{BABAR1,BABAR2} The $\eta_b$ meson
shows up as a peak in the photon energy spectrum of the $\Upsilon\to
\gamma\eta_b$ transitions. Despite considerable background from
$\Upsilon\to \gamma\chi_b$ and
 $e^+e^-\to \gamma \Upsilon$  processes, the peak energy
can be measured with  rather high precision.
Together  with very high accuracy of  the $\Upsilon$ spectroscopy,
this allows  for  the determination of  $\eta_b$ mass $M(\eta_b)$
with only a few MeV error. The analysis of the $\Upsilon(3S)$ decays
gives $M(\eta_b)={9388.9}^{+ 3.1}_{-2.3}\,
({\rm stat})\pm 2.7\,{(\rm syst)} ~{\rm MeV}$,\cite{BABAR1} while
$\Upsilon(2S)$ data give $M(\eta_b)={9392.9}^{+ 4.6}_{-4.8}\,
({\rm stat})\pm 1.9\,{(\rm syst)} ~{\rm MeV}$.\cite{BABAR2}
Thus an accurate prediction of  $M(\eta_b)$ is
a big challenge and a test for the QCD theory of heavy quarkonium.
Due to  a very small experimental uncertainty
of the $\Upsilon(1S)$ mass, the problem can be  reduced to the
calculation of the hyperfine splitting (HFS) $E_{\rm
hfs}=M(\Upsilon(1S))-M(\eta_b)$. This quantity  is very sensitive to
$\alpha_s$ and could become a competitive source for the determination of the
strong coupling constant. In this paper we briefly review  the advances
and problems in the QCD theory of bottomonium HFS. We  consider only the approaches entirely
based on the first principles of QCD, leaving aside numerous
semi-phenomenological models.

\section{Bottomonium Hyperfine Splitting in QCD}

\begin{figure}[t]
\begin{center}
\epsfig{figure=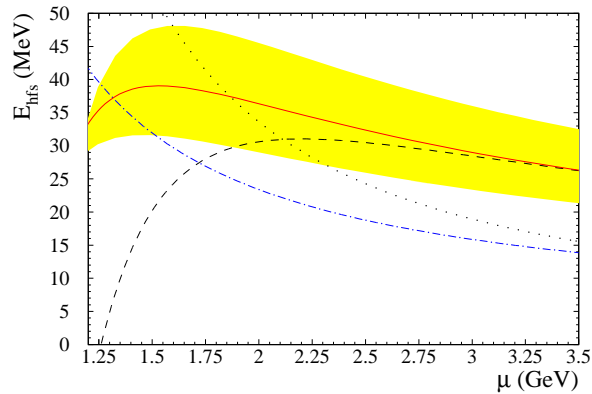,height=6cm}
\end{center}
\caption{HFS of 1S bottomonium as a function of the
renormalization scale $\mu$ in the LO (dotted line), NLO (dashed line), LL
(dot-dashed line), and NLL (solid line) approximations. For the NLL
result, the band reflects the errors due to $\alpha_s(M_Z)=0.118\pm
0.003$.
\label{fig1}}
\end{figure}

Systematic  perturbative analysis  of the
heavy quarkonium bound states
is based on the effective field theory  of
(potential) nonrelativistic QCD, or (p)NRQCD.\cite{CasLep,PinSot1}
A recent major  breakthrough in the high-order calculations
of the heavy quarkonium properties is related to the use of
dimensional regularization \cite{PinSot2} and the threshold expansion \cite{BenSmi}
within the effective field theory framework.\cite{PenSte,KniPen1}
The bottomonium spectrum has been computed to
${\cal O}(m_b\alpha_s^5)$, which includes the ${\cal O}(\alpha_s)$
next-to-leading order (NLO)
correction to  the HFS.
The corresponding result for an arbitrary  principal quantum number
is given in
Refs.\cite{PenSte,PSS} in a closed analytical form. For the ground state it reads
\begin{eqnarray}
 {E_{\rm hfs}^{NLO}}& = &{ C_F^4 \alpha_s^4m_b\over 3}
\left[
1 + {{\alpha_s \over \pi}}\,\left(
  \frac{7\,{C_A}\,}{4}
{\ln\left({C_F\alpha_s}\right)}
 -{{C_F}\over 2}+ \frac{2\pi^2-26}{9} {n_f}\,{T_F}
+\frac{3-3\ln\,2}{2}T_F
\right.\right.
\nonumber\\
&+&\left.\left.\frac{
122-11\pi^2}{18}{C_A}\right)\right]  \approx
{E_{\rm hfs}^{LO}}\left[
1 + {{\alpha_s}}\,\left(
  1.67\,
{\ln\left({\alpha_s}\right)}
 +0.61
\right)\right]\,,
\label{eq:pqcd}
\end{eqnarray}
where
$C_F=(N_c^2-1)/(2N_c)$, $C_A=N_c=3$, $n_l=4$, and
$\alpha_s$ is renormalized in the $\overline{\rm MS}$ scheme
at the scale $\mu=C_F\alpha_sm_b$. A logarithmically enhanced term in
Eq.~(\ref{eq:pqcd}) is characteristic to the multiscale
dynamics of the nonrelativistic bound states.\cite{KniPen2}
Such terms can be resummed to all orders through the renormalization group
analysis of pNRQCD, or the {\it nonrelativistic renormalization group}
(NRG) \cite{Pin,PPSS} (see also  Ref.\cite{LMR}).
The renormalization-group-improved expression for the bottomonium HFS
is available to  the next-to-leading logarithmic (NLL) approximation,
which sums up all the corrections of the form $\alpha_s^n\ln^{n-1}\alpha_s$.\cite{KPPSS}
The corresponding analytical expression is too lengthy to
be presented here. The result of the numerical analysis is given in Fig.~\ref{fig1}.
The logarithmic expansion shows nice convergence and weak
scale dependence  at the physical  scale of the inverse Bohr radius
$\mu\sim \alpha_sm_b$.  This suggests a small
uncertainty due to
uncalculated higher-order terms.  At the same time the nonperturbative contribution to the HFS
is difficult to estimate. In principle it can be investigated by the method of vacuum
condensate expansion.\cite{VolLeu}  The resulting series, however, does not
converge well and suffers from large numerical
uncertainties.\cite{TitYnd2}  On the other hand, the nonperturbative contribution
is suppressed at least by the second power of the heavy quark velocity $v\sim \alpha_s$.
Hence it is beyond the accuracy of the NLL  approximation and  should be added to the
errors.  In the  charmonium system, where the nonperturbative
effects are supposed to be much more important, the NLL
approximation  gives the central value
$M(J/\psi)-M(\eta_c)=104$~MeV,\cite{KPPSS} which is in a very good
agreement with the experimental value  $117.7\pm 1.3$~MeV. This suggests that the
nonperturbative contribution to the  bottomonium HFS  is  likely
to be small as well. A detailed discussion of the  uncertainties
of the NLL result can be found in Ref.\cite{KPPSS} The final numerical
prediction for the bottomonium HFS based on perturbative QCD reads
\begin{equation}
E^{\rm QCD}_{\rm hfs}=39 \pm 11\,{(\rm th)} \,{}^{+9}_{-8}\,
(\delta\alpha_s)~{\rm MeV}\,,
\label{eq:res}
\end{equation}
where ``th'' stands for the errors due to the high-order perturbative
corrections and nonperturbative effects,
whereas ``$\delta\alpha_s$'' stands for the uncertainty in
$\alpha_s(M_Z)=0.118\pm0.003$.

The problem of  proper description of the
nonperturbative dynamics of  strong interactions  at long distance
is naturally solved by the lattice simulations of  QCD.
A systematic analysis of the bottomonium HFS within the unquenched lattice
NRQCD  predicts \cite{Gra}
\begin{equation}
E^{\rm lat}_{\rm hfs}={61}\pm 14~{\rm MeV},
\label{eq:lat}
\end{equation}
which has somewhat larger central value than  Eq.~(\ref{eq:res}), but
agrees with the perturbative result within the error bars.

\section{Discussion}
The estimate~(\ref{eq:pqcd}) based on the
perturbative QCD undershoots
the experimentally measured  values
\begin{equation}
E^{\rm exp}_{\rm hfs}=
71.4^{+2.3}_{-3.1}\,(stat) \pm 2.7 \,(syst)~{\rm MeV},\mbox{\cite{BABAR1}}
\qquad E^{\rm exp}_{\rm hfs}= 67.4^{+4.8}_{-4.6}\,(stat) \pm 2.0\,
(syst)~{\rm MeV},\mbox{\cite{BABAR2} }
\end{equation}
by  about  two standard deviations. This discrepancy is rather unexpected
and difficult to explain if one takes into account the very successful perturbative
description of the HFS in charmonium. At the same time the  prediction
of the lattice QCD   apparently agrees with the experimental data. This fact, however, should be taken with
great care. Indeed, the lattice simulation \cite{Gra} uses a finite lattice
spacing $a\sim (\alpha_sm_b)^{-1}$.
It is determined by fitting the bottomonium spectrum, which is mostly sensitive
to the soft momentum scale $\alpha_sm_b$.
At the same time the HFS gets a significant  contribution from the hard momentum
scale of the heavy quark mass
through the radiative corrections.  In the lattice NRQCD framework this contribution should be included into the
Wilson coefficient of the spin-flip operator in the effective Hamiltonian, which
is  neglected in Eq.~(\ref{eq:lat}).
The one-loop Wilson coefficient contains a large logarithm of the form $\ln\left(am_b\right)$. It is in one-to-one correspondence
with the logarithmic term of Eq.~(\ref{eq:pqcd}) and results in an additional contribution to the
HFS
\begin{equation}
\delta^{\rm hard}E_{\rm hfs}= - {{\alpha_s \over \pi}}
  \frac{7\,{C_A}\,}{4}
{\ln\left(am_b\right)}E_{\rm hfs}\approx -20~{\rm MeV},
\end{equation}
which  brings the lattice estimate~(\ref{eq:lat})  in a perfect agreement with
the perturbative result~(\ref{eq:pqcd}). Thus,
no definite conclusion on the accuracy  of the lattice QCD
predictions for the bottomonium HFS can be made at the moment  and  further
theoretical study is necessary. In particular one has to compute the Wilson
coefficient of the spin-flip operator perturbatively  in the lattice
regularization beyond the logarithmic approximation.

To summarize, with the precise experimental data
now at hand,  the bottomonium HFS becomes one of the
most interesting hadronic  systems to apply and to test the QCD theory
of strong interactions.  A significant
discrepancy between the prediction based on perturbative QCD and the
experimentally measured HFS is intriguing and requires
further analysis.

\section*{Acknowledgments}
This work  is supported  by the
Alberta Ingenuity foundation
and NSERC.

\end{document}